\documentstyle[mathrsfs,amsmath,amsfonts,epic,eepic,epsfig]{elsart}

\begin{document}

\begin{frontmatter}

\title{Neutrino oscillations with three flavors in matter:
Applications to neutrinos traversing the Earth}

\author{Tommy Ohlsson\thanksref{emailtommy}},
\author{H{\aa}kan Snellman\thanksref{emailsnell}}
\address{Division of Mathematical Physics, Theoretical Physics,
Department of Physics, Royal Institute of Technology, SE-100~44
Stockholm, Sweden}

\thanks[emailtommy]{E-mail address: tommy@theophys.kth.se}
\thanks[emailsnell]{E-mail address: snell@theophys.kth.se}

\date{Received 15 December 1999}

\begin{abstract}
Analytic formulas are presented for three flavor neutrino oscillations
in matter in the plane wave approximation. We calculate in particular
the time evolution operator in both mass and flavor bases. We also
find the transition probabilities expressed as functions of the vacuum
mass squared differences, the vacuum mixing angles, and the matter
density parameter. The application of this to neutrino oscillations
for both atmospheric and long baseline neutrinos in a mantle-core-mantle 
step function model of the Earth's matter density profile is discussed.

\end{abstract}

\end{frontmatter}

\section{Introduction}
\label{sec:intro}

As an accumulating amount of data on neutrino oscillations is becoming
accessible, it is of interest to study three flavor neutrino oscillations.
Here we would like to give analytic expressions for the neutrino oscillation
probabilities in presence of matter expressed in the mixing matrix
elements and the neutrino energies or masses, {\it i.e.},
incorporating the so called Mikheyev--Smirnov--Wolfenstein (MSW)
effect \cite{mikh85,wolf78}. These probability formulas are of interest for
the solar neutrino problem, the atmospheric neutrino problem, and the
long baseline (LBL) neutrino experiments. We here apply the
formalism to neutrinos traversing the Earth in a mantle-core-mantle
step function model of the Earth's matter density profile.

We will assume that $CP$ nonconservation is negligible at the
present level of experimental accuracy \cite{dick99}. Thus, the mixing
matrix for the neutrinos is real.

Previous work on models for three flavor neutrino oscillations in
matter includes works of Barger et al. \cite{barg80}, Kim and
Sze \cite{kim87}, and Zaglauer and Schwarzer
\cite{zagl88}. Our method is different from all these approaches and their
parameterizations also differ slightly from ours. In particular, we
calculate the time evolution operator and do not use the
auxiliary matter mixing angles.

Approximate solutions for three flavor neutrino
oscillations in matter have been presented by Kuo and Pantaleone
\cite{kuo86} and Joshipura and Murthy \cite{josh88}. Approximate
treatments have also been done by Toshev and Petcov \cite{tosh87}.
D'Olivo and Oteo have made contributions by using an approximative
Magnus expansion for the time evolution operator \cite{oliv96}.
Extensive numerical investigations for matter enhanced three flavor
oscillations have been made by Fogli et al. \cite{fogl94}.
Studies of neutrino oscillations in the Earth has been performed by
several authors \cite{nico88,kras88,giun98,liu98,petc98,harr99,freu99}.

\section{The evolution operator}

Let the flavor state basis and mass eigenstate basis be given by
${\cal H}_{f} \equiv \{ \vert \nu_\alpha \rangle
\}_{\alpha=e,\mu,\tau}$ and ${\cal H}_{m} \equiv \{ \vert \nu_a \rangle
\}_{a=1}^3$, respectively. Then the flavor
states $\vert \nu_\alpha \rangle \in {\cal H}_{f}$ can be obtained as
superpositions of the mass eigenstates $\vert \nu_a \rangle \in {\cal
H}_{m}$, or vice versa. The bases ${\cal H}_f$ and ${\cal
H}_m$ are of course just two different representations of the same
Hilbert space ${\cal H}$.

In the present analysis, we will use the plane wave approximation to
describe neutrino oscillations. In this approximation, a neutrino
flavor state $|\nu_{\alpha}\rangle$ is a linear combination of
neutrino mass eigenstates $|\nu_{a} \rangle$ such that \cite{kim93}
\begin{equation}
\vert \nu_\alpha \rangle = \sum_{a=1}^3 U^\ast_{\alpha a} \vert \nu_a
\rangle,
\end{equation}
where $\alpha = e,\mu,\tau$. In what follows, we will
use the short-hand notations $\vert
\alpha \rangle \equiv \vert \nu_\alpha \rangle$ and $\vert a \rangle
\equiv \vert \nu_a \rangle$ for the flavor states and the mass
eigenstates, respectively.

For the components of a state $\psi$ in the flavor state basis and mass
eigenstate basis, respectively, they are related to each other by
\begin{equation}
\psi_f = U \psi_m,
\label{eq:flavor1}
\end{equation}
where
$$
\psi_f = \left( \psi_\alpha \right) = \left( \begin{array}{c} \psi_e
\\ \psi_\mu \\ \psi_\tau
\end{array} \right) \in {\cal H}_f \quad \mbox{and} \quad \psi_m =
\left( \psi_a \right) = \left(
\begin{array}{c} \psi_1 \\ \psi_2 \\ \psi_3 \end{array} \right) \in
{\cal H}_m.
$$

A convenient parameterization for $U = U(\theta_1,\theta_2,\theta_3)$
is given by \cite{caso98}
\begin{equation}
U = \left( \begin{array}{ccc} C_2 C_3 & S_3 C_2 & S_2 \\ - S_3 C_1 -
S_1 S_2 C_3 & C_1 C_3 - S_1 S_2 S_3 & S_1 C_2 \\ S_1 S_3 - S_2
C_1 C_3 & - S_1 C_3 - S_2 S_3 C_1 & C_1 C_2 \end{array} \right),
\end{equation}
where $S_i \equiv \sin \theta_i$ and $C_i \equiv \cos \theta_i$ for $i
= 1,2,3$. This is the standard representation of the
mixing matrix. The quantities $\theta_i$, where $i = 1,2,3$, are
the vacuum mixing angles. Since we have put the $CP$
phase equal to zero in the mixing matrix, this means that $U^\ast_{\alpha
a} = U_{\alpha a}$ for $\alpha = e, \mu, \tau$ and $a = 1,2,3$.

In the mass eigenstate basis, the Hamiltonian ${\mathscr H}$ for the
propagation of
the neutrinos in vacuum is diagonal and given by
\begin{equation}
H_m = \left( \begin{array}{ccc} E_{1} & 0 & 0 \\ 0 & E_{2} & 0 \\ 0 &
 0 & E_{3} \end{array} \right),
\end{equation}
where $E_a = \sqrt{m_a^2 + {\bf p}^2}$, $a = 1,2,3$, are the energies
of the neutrino mass eigenstates $\vert a \rangle$, $a = 1,2,3$ with
masses $m_a$, $a = 1,2,3$. We will assume ${\bf p}$ to be the
same for all mass eigenstates.

When neutrinos propagate in matter, there is an additional term coming
from the presence of electrons in matter \cite{wolf78}. This term
is diagonal in the flavor state basis and is given by
\begin{equation}
V_{f} = \left( \begin{array}{ccc} A & 0 & 0 \\ 0 & 0 & 0 \\ 0 & 0 &
 0\end{array} \right),
\end{equation}
where
$$
A = \pm \sqrt{2}G_F N_{e} \simeq \pm \frac{1}{\sqrt{2}} G_F
\frac{1}{m_N} \rho
$$
is the matter density parameter.
Here $G_F$ is the Fermi weak coupling constant, $N_{e}$ is the electron
density, $m_N$ is the nucleon mass, and $\rho$ is the matter density.
The sign depends on whether we deal with neutrinos~($+$) or
antineutrinos~($-$). We will assume that the electron density $N_e$
(or the matter density $\rho$) is constant throughout the matter in
which the neutrinos are propagating. In the mass basis, this piece of
the Hamiltonian is $V_{m}=U^{-1}V_{f}U$, where again $U$ is the mixing matrix.

The unitary transformation that leads from the initial state $\psi_{f}(0)$
in flavor basis at time $t=0$ of production of the neutrino, to the state
of the same neutrino $\psi_{f}(t)$ at the detector at time $t$ is given by
the operator $U_{f}(t)\equiv U_{f}(t,0)$, where $U_{f}(t_{2},t_{1})$ is the
time evolution operator from time $t_{1}$ to time $t_{2}$ in flavor basis.
This operator can be formally written as $U_{f}(t)=e^{-iH_{f}t}$. When the
neutrinos are propagating through vacuum, the Hamiltonian in flavor basis
is $H_{f}= UH_{m}U^{-1}$. In this case, the exponentiation of $H_{f}$ can
be performed easily: $U_{f}(t)=e^{-iH_f t} = Ue^{-iH_{m}t}U^{-1}$, and
the result can be expressed in closed form. In the case when the neutrinos
propagate through matter, the Hamiltonian is not diagonal in either the
mass eigenstate basis or the flavor state basis, and we have to calculate the
evolution operator $U_{f}(t)$ or $U_f(L) \equiv e^{-i {\mathscr H}_f L} = U
e^{-i {\mathscr H}_m L} U^{-1}$ if we set $t=L$.

To do so it is convenient to introduce the traceless matrix $T$
defined by $T \equiv {\mathscr H}_{m} - ({\rm tr\,} {\mathscr H}_{m}) I/3$.
The trace of the Hamiltonian in the mass basis
${\mathscr H}_m \equiv H_m + U^{-1} V_f U$ is
${\rm tr\,} {\mathscr H}_m = E_1 + E_2 + E_3 + A$,
and the matrix $T$ can then be written as
\begin{eqnarray}
T &=& (T_{ab}) \nonumber\\
&=& \left( \begin{array}{ccc} A U_{e1}^2 - \tfrac{1}{3} A +
\tfrac{1}{3} \left( E_{12} + E_{13} \right) & A U_{e1} U_{e2} &A
U_{e1} U_{e3} \\ A U_{e1} U_{e2} & A U_{e2}^2 - \tfrac{1}{3} A +
\tfrac{1}{3} \left( E_{21} + E_{23} \right) & A U_{e2} U_{e3} \\ A
U_{e1} U_{e3} & A U_{e2} U_{e3} & A U_{e3}^2 - \tfrac{1}{3} A +
\tfrac{1}{3} \left( E_{31} + E_{32} \right)
\end{array} \right), \nonumber\\
\label{eq:T}
\end{eqnarray}
where $E_{ab} \equiv E_a - E_b$. Of the six quantities
$E_{ab}$, where $a,b=1,2,3$ and $a \neq b$, only two are linearly independent,
since the $E_{ab}$'s fulfill the relations $E_{ba} = - E_{ab}$ and
$E_{12} + E_{23} + E_{31} = 0$.\footnote{Later, we will use the usual
(vacuum) mass squared differences $\Delta m^2_{21}$ and $\Delta
m^2_{32}$ instead of $E_{21}$ and $E_{32}$, which are related to each
other by $\Delta m_{21}^2 = 2 E_\nu E_{21}$ and $\Delta m_{32}^2 = 2 E_\nu
E_{32}$, where $E_\nu$ is the neutrino energy.}
Using Eq.~(\ref{eq:T}), the evolution operator in the mass basis can
be written as
\begin{equation}
U_m(L) \equiv e^{-i {\mathscr H}_m L} = \phi e^{-i L T},
\label{eq:UmL}
\end{equation}
where $\phi \equiv e^{-i L {\rm tr \,}{\mathscr H}_{m}/3}$.

The flavor states can be expressed either as linear combinations of the vacuum
mass eigenstates ($A = 0$) in the basis ${\cal H}_{m}$ as in
Eq.~(\ref{eq:flavor1}) or
the matter mass eigenstates ($A \neq 0$) in the basis ${\cal H}_{M}$.
In the latter case, the corresponding components are related to each other as
\begin{equation}
\psi_f = U^{M} \psi_M,
\end{equation}
where $U^{M} = U^{M}(\theta_1^M, \theta_2^M, \theta_3^M)$ is the unitary mixing
matrix for matter and $\theta^M_i$,
$i = 1,2,3$, are the (auxiliary) matter mixing angles.

Combining these expressions for the flavor state components, one obtains the
following relation between the two different sets of mass eigenstate
components
\begin{equation}
\psi_M = R \psi_m,
\end{equation}
where
$$
R \equiv (U^M)^{-1} U.
$$
The matrix $R$ is, of course, a unitary matrix (even orthogonal,
since $U$ and $U^M$ are real). This means that the matter mixing
matrix can be expressed in the vacuum mixing matrix as
\begin{equation}
U^{M}(\theta_1^M, \theta_2^M, \theta_3^M) =
U(\theta_1,\theta_2,\theta_3) R^{-1}.
\label{eq:UUR}
\end{equation}

The relations between the different bases are depicted in the
following diagram:

\begin{center}
\begin{picture}(8,4)
\put(1,3){$\psi_m \in {\cal H}_m$}
\put(3,3.1){\vector(1,0){1.75}}
\put(5.25,3){${\cal H}_f \ni \psi_f$}
\put(2.25,2.5){\vector(0,-1){1.5}}
\put(1,0.25){$\psi_{M} \in {\cal H}_M$}
\put(3,1){\vector(4,3){1.9}}
\put(3.75,3.5){$U$}
\put(1.6,1.65){$R$}
\put(4.15,1.15){$U^{M}$}
\end{picture}
\end{center}

From this diagram one readily obtains
$$
{\mathscr H}_M \equiv R {\mathscr H}_m R^{-1} \quad \mbox{or}
\quad {\mathscr H}_M \equiv R T R^{-1} + \tfrac{1}{3} ({\rm tr \,}
{\mathscr H}_m) I,
$$
where ${\mathscr H}_{M}$ is the Hamiltonian in matter, which is diagonal
in the basis ${\cal H}_{M}$.

Due to the invariance of the trace, we have
\begin{equation}
T_{M} \equiv {\mathscr H}_{M} -\tfrac{1}{3}({\rm tr\,}{\mathscr
H}_{M})I=RTR^{-1}.
\label{eq:tm}
\end{equation}
and $T_{M}$ is a diagonal matrix with elements $\lambda_{a}$,
$a=1,2,3$, the eigenvalues of $T$. This implies that
\begin{equation}
e^{-iLT_{M}}=Re^{-iLT}R^{-1}.
\label{eq:exptm}
\end{equation}

Now, Cayley--Hamilton's theorem implies that, since $T$ is a
$3 \times 3$ matrix, the infinite series defining $e^{-iLT}$
can be written as a second order polynomial in $T$ \cite{ohls99}:
\begin{equation}
e^{-i LT} = a_0 I - i LT a_1 - L^2 T^2 a_2,
\label{eq:fin_series}
\end{equation}
where $a_0$, $a_1$, and $a_2$ are coefficients to be
determined.
Inserting Eq.~({\ref{eq:fin_series}) into Eq.~(\ref{eq:exptm}) and using
Eq.~(\ref{eq:tm}) gives a linear system of three equations that will
determine the coefficients $a_0$, $a_1$, and $a_2$:
\begin{equation}
\left\{ \begin{array}{l} e^{-i L \lambda_1} = a_0 - i L \lambda_1 a_1
- L^2 \lambda_1^2 a_2 \\
e^{-i L \lambda_2} = a_0 - i L \lambda_2 a_1 - L^2 \lambda_2^2 a_2 \\
e^{-i L \lambda_3} = a_0 - i L \lambda_3 a_1 - L^2 \lambda_3^2 a_2
\end{array} \right.,
\label{eq:eqsys}
\end{equation}
where $\lambda_a$, $a=1,2,3$, are the diagonal elements of $T_{M}$, or,
equivalently, the eigenvalues of $T$, {\it i.e.}, the
solutions to the characteristic equation
\begin{equation}
\lambda^3 + c_2 \lambda^2 + c_1 \lambda + c_0 = 0
\end{equation}
with
\begin{eqnarray}
c_{2} &=& - {\rm tr \,} T = 0,\\
c_{1} &=& \det T \, {\rm tr \,} T^{-1},\\
c_{0} &=& - \det T.
\label{eq:coeff}
\end{eqnarray}
The coefficients $c_0$, $c_1$, and $c_2$ are all real and the
eigenvalues $\lambda_{a}$, $a=1,2,3$, can be expressed in closed form
in terms of these \cite{ohls99}.

When the system of equations (\ref{eq:eqsys}) is solved for the
$a_{i}$'s, we obtain
\begin{equation}
e^{-iLT}= \sum_{a=1}^{3} e^{-iL\lambda_{a}}
\frac{1}{3\lambda_{a}^{2}+c_{1}} \left[ (\lambda_{a}^{2}+c_{1})I
+\lambda_{a}T+T^{2} \right].
\label{eq:exptm1}
\end{equation}
From Eqs.~(\ref{eq:UmL}) and (\ref{eq:exptm1}) we then obtain
\begin{equation}
U_m(L) = e^{-i {\mathscr H}_m L} = \phi e^{-i L T} = \phi \sum_{a=1}^3
e^{-i L \lambda_a} \frac{1}{3\lambda_a^2+c_1} \left[ (\lambda_a^2 +
c_1)I + \lambda_a T + T^2 \right].
\label{eq:eiH_ml_fin_sum}
\end{equation}
The evolution operator for the neutrinos in flavor basis is thus given by
\begin{equation}
U_{f}(L) = e^{-i {\mathscr H}_{f} L} = U e^{-i {\mathscr H}_{m} L}
U^{-1} = \phi \sum_{a=1}^3 e^{-i L \lambda_a}
\frac{1}{3\lambda_a^2+c_1} \left[ (\lambda_a^2 + c_1)I + \lambda_a \tilde{T} +
\tilde{T}^2 \right],
\label{eq:evol}
\end{equation}
where $\tilde{T} \equiv U T U^{-1}$. 
Formula~(\ref{eq:evol}) is the final expression for the evolution
operator. It expresses the time (or $L$) evolution directly in term
of the mass squared differences and the vacuum mixing angles without
introducing the auxiliary matter mixing angles.

\section{Transition probabilities}
\label{sec:prob}

The probability amplitude is defined as
\begin{equation}
A_{\alpha\beta} \equiv \langle \beta \vert U_f(L) \vert \alpha
\rangle, \quad \alpha,\beta=e,\mu,\tau.
\label{eq:ampl}
\end{equation}
Inserting Eq.~(\ref{eq:evol}) into Eq.~(\ref{eq:ampl}) gives
\begin{equation}
A_{\alpha\beta} = \phi \sum_{a=1}^3 e^{-i L \lambda_a}
\frac{(\lambda_a^2 + c_1)
\delta_{\alpha\beta} + \lambda_a \tilde{T}_{\alpha\beta} +
(\tilde{T}^2)_{\alpha\beta}}{3\lambda_a^2+c_1},
\end{equation}
where $\delta_{\alpha\beta}$ is Kronecker's delta.
Note that $\tilde{T}_{\alpha\beta} =
\tilde{T}_{\beta\alpha}$ and $(\tilde{T}^2)_{\alpha\beta} =
(\tilde{T}^2)_{\beta\alpha}$.

Inserting $L=0$ into Eq.~(\ref{eq:evol}) yields
\begin{equation}
\delta_{\alpha \beta} = \sum_{a=1}^3 \frac{(\lambda_a^2 +
c_1) \delta_{\alpha\beta} + \lambda_a \tilde{T}_{\alpha\beta} +
(\tilde{T}^2)_{\alpha\beta}}{3 \lambda_a^2 + c_1}.
\end{equation}
Hence, the transition probabilities in matter are
\begin{eqnarray}
P_{\alpha\beta} = \vert A_{\alpha\beta} \vert^2 &=& \delta_{\alpha\beta}
- 4 \; \underset{a < b}{\sum_{a=1}^3
\sum_{b=1}^3} \frac{(\lambda_a^2 + c_1) \delta_{\alpha\beta} +
\lambda_a \tilde{T}_{\alpha\beta} +
(\tilde{T}^2)_{\alpha\beta}}{3 \lambda_a^2 + c_1} \nonumber\\
&\times& \frac{(\lambda_b^2 + c_1) \delta_{\alpha\beta} + \lambda_b
\tilde{T}_{\alpha\beta} + (\tilde{T}^2)_{\alpha\beta}}{3 \lambda_b^2 + c_1}
\sin^{2}\tilde{x}_{ab}, \quad \alpha,\beta = e,\mu,\tau, \nonumber\\
\label{eq:Pab_2}
\end{eqnarray}
where $\tilde{x}_{ab} \equiv (\lambda_a - \lambda_b)L/2$.

From unitarity, there are only three independent transition
probabilities, since the other three can be obtained from them, {\it
i.e.}, from the equations
\begin{eqnarray}
&& P_{ee} + P_{e\mu} + P_{e\tau} = 1, \label{eq:pe1}\\
&& P_{\mu e} + P_{\mu\mu} + P_{\mu\tau} = 1, \label{eq:pe2}\\
&& P_{\tau e} + P_{\tau\mu} + P_{\tau\tau} = 1. \label{eq:pe3}
\end{eqnarray}
Note that $P_{e\mu} = P_{\mu e}$, $P_{e\tau} = P_{\tau e}$, and
$P_{\mu\tau} = P_{\tau\mu}$. If we choose $P_{e\mu}$, $P_{e\tau}$,
and $P_{\mu\tau}$ as the three independent ones, we thus have for $\alpha
\neq \beta$

\begin{equation}
P_{\alpha\beta}=
- 4 \; \underset{a < b}{\sum_{a=1}^3
\sum_{b=1}^3} \frac{\lambda_a \tilde{T}_{\alpha\beta} +
(\tilde{T}^2)_{\alpha\beta}}{3 \lambda_a^2 + c_1}
\frac{\lambda_b \tilde{T}_{\alpha\beta} +
(\tilde{T}^2)_{\alpha\beta}}{3 \lambda_b^2 + c_1}
\sin^{2}\tilde{x}_{ab}.
\label{eq:Pab_n1}
\end{equation}

\section{Applications and discussion}
\label{sec:disc}

The main results of our analysis are given by the time evolution
operator for the neutrinos when passing through matter with constant matter
density in Eq.~(\ref{eq:evol}) and the expression for the transition
probabilities in Eqs.~(\ref{eq:Pab_2}) and (\ref{eq:Pab_n1}),
expressed as finite sums of elementary functions in the
elements of $\tilde{T}$.

In our treatment the auxiliary mixing angles in matter, $\theta^M_i$, play no
independent role and are not really needed.

When the neutrinos travel through a series of matter densities with
matter density parameters $A_{1},\ldots,A_{n}$ and thicknesses
$L_{1},\ldots,L_{n}$, the total evolution operator is simply given by
\begin{equation}
U_{f}(L)=\prod_{i=1}^{n}U_{f}(L_{i}) = U_{f}(L_{n}) \ldots
U_{f}(L_{1}),
\label{eq:prod}
\end{equation}
where $L \equiv \sum_{i=1}^n L_{i}$ and $U_{f}(L_{i})$ is calculated
for $A=A_{i}$. Equation~(\ref{eq:prod}) gives a simple and fast
algorithm to obtain the total evolution operator instead of using
numerical integration.

As an application to show the usefulness of our derived formulas, we
have calculated the transition probabilities $P_{\alpha\beta}$ for
neutrino oscillations in a mantle-core-mantle step function model
simulating the Earth's matter density profile.
Let $R \simeq 6371$ km be the radius of the Earth and $r \simeq 3486$ km be the
radius of the core. The thickness of the mantle is then $R-r \simeq 2885$ km,
with matter density parameter $A_{1} \simeq 1.70 \cdot 10^{-13}$ eV ($\rho_1
\simeq 4.5$ ${\rm g/cm^3}$), whereas the matter
density parameter of the core is $A_{2} \simeq 4.35 \cdot 10^{-13}$ eV
($\rho_2 \simeq 11.5$ ${\rm g/cm^3}$).

Neutrinos traversing the Earth towards a detector close
to the surface of the Earth, pass through the matter densities $A_{1}, A_{2},
A_{1}$ of thicknesses $L_{1}, L_{2}, L_{1}$ where the distances
$L_{i}$, $i=1,2$, are functions of the nadir angle $h$, where $h \equiv
180^\circ-\theta_{z}$; $\theta_{z}$ being the zenith angle. As $h$ varies
from $0$ to $90^\circ$, the cord $L = L(h)$ of the neutrino passage through
the Earth becomes shorter and shorter. At an angle larger than $h_{0} =
\arcsin (r/R) \simeq 33.17^{\circ}$, the
distance $L_{2}=0$, and the neutrinos no longer traverse the core.

For $0\leq h\leq h_{0}$ the distances $L_{1}$ and $L_{2}$ are given by
\begin{eqnarray}
L_{1} &=& R \left( \cos h
-\sqrt{\left( \frac{r}{R} \right) ^{2}-\sin^{2}h} \right), \\
L_{2} &=& 2R\sqrt{\left(\frac{r}{R}\right)^{2}-\sin^{2}h}, \\
L &=& 2 L_1 + L_2.
\end{eqnarray}
For $h_{0}\leq h \leq 90^\circ$:
\begin{equation}
L = 2R\cos h.
\end{equation}

The mass squared differences ($\Delta M^{2} \equiv \Delta m_{32}^2$
and $\Delta m^{2} \equiv \Delta m_{21}^2$) and
the vacuum mixing angles ($\theta_{1},\theta_{2},\theta_{3}$) used here
are chosen to correspond to those obtained from analyses of various neutrino
oscillation data. For our illustration we have taken the following
parameter values
$$
\Delta M^{2} = 3.2 \cdot 10^{-3} {\rm eV}^{2}, \quad
\Delta m^{2} = 0, \Delta M^{2}/10, \quad
\theta_{1} = 45^{\circ}, \quad
\theta_{2} = 5^{\circ}, \quad
\theta_{3} = 45^{\circ}.
$$
The values of $\Delta M^{2}$ and $\theta_{1}$ are governed by
atmospheric neutrino data \cite{scho99} and the values of $\Delta m^{2}$ and
$\theta_{3}$ by solar neutrino data \cite{bahc98}. The value of $\theta_{2}$ is
below the CHOOZ upper bound, which is $\sin^2 2\theta_2 = 0.10$ \cite{apol98}.
These choices are the most optimistic ones for obtaining any effects
in LBL experiments from the sub-leading $\Delta m^{2}$ scale
\cite{barg99}. We should mention though, that these data are taken from
two flavor model analyses.

The probabilities corresponding to various situations in this
scenario are illustrated in Figs.~\ref{fig:1} - \ref{fig:4}. These figures
correspond to the physically measurable quantities.
From the figures one can see that there are several resonance phenomena
superimposed on each other.

Figures~\ref{fig:1} a) - \ref{fig:1} d) show the results for the case
of $h=0$, one mass squared difference equal to $3.2 \cdot 10^{-3}$ ${\rm
eV}^{2}$ \cite{scho99}, and the other
one equal to 0. In Fig.~\ref{fig:1} a) we have included for comparison
the corresponding result of a two flavor model. Figures~\ref{fig:1}
b) - \ref{fig:1} d) show the probabilities $P_{ee}$, $P_{e\mu}$, and
$P_{\mu\tau}$, respectively, in the three flavor model. The
other probabilities can be obtained by unitarity from these three
using Eqs.~(\ref{eq:pe1}) - (\ref{eq:pe3}). We observe that
although the survival probability $P_{ee}$ is the same for both cases,
the transition probability $P_{e\mu}$ is only half of that in the two
flavor model. In a two flavor model, only one transition probability
is needed, since if $P_{ee}$ is given and $P_{\mu e} = P_{e\mu}$, the
other two follow from $P_{e\mu} = 1 - P_{ee}$ and $P_{\mu\mu} = 1 -
P_{e\mu} = P_{ee}$.

In Figs.~\ref{fig:2} a) - d) the second mass squared difference is
chosen to be $1/10$ of the first one. This affects the
disappearance rate for low energies in the three flavor case, as can be
seen from Fig.~\ref{fig:2} b). It also modifies the appearance rates
$P_{e\mu}$ and $P_{\mu\tau}$ in Figs.~\ref{fig:2} c) and \ref{fig:2} d),
respectively. Figure~\ref{fig:2} a) is again the same as Fig.~\ref{fig:1} a).

When the small mass squared difference is much smaller than the large
mass squared difference, then the two flavor model coincides with the
three flavor model. This is not the case when the small mass squared
difference is of comparable size to the large mass squared
difference. See Figs.~\ref{fig:1} a), \ref{fig:1} b), \ref{fig:2} a),
and \ref{fig:2} b) for a comparison.

Figure~\ref{fig:3} shows a contour plot of $P_{ee}$ as a function of
the matter density parameter of the core $A_{\rm core}$ (as it changes
from the value of the mantle $A_1$ up to its full value $A_2$) and the
neutrino energy $E_{\nu}$. The dark areas correspond to regions of
large conversion probability for the assumed parameter values. When
the small mass squared difference decreases, the structure below the
main conversion valley goes away. Furthermore, the conversion regions
above the main conversion valley are obviously due to the core (either
purely to the core or to interference effects between the core and the
mantle).

Similarly, Fig.~\ref{fig:4} shows a contour plot of $P_{ee}$ as a
function of the nadir angle $h$ and the neutrino energy $E_{\nu}$. As in the
previous figure, the dark areas correspond
to regions of large conversion probability for the assumed parameter
values. At $h = h_0$, the neutrinos no longer traverse
the core, but only the mantle. Again, when the small mass squared difference
drops, the oscillations in the bottom of the plot (sub-GeV region) go away and
no conversion takes place, {\it i.e.}, $P_{ee} \simeq 1$ in this
region. In the limit when the nadir angle goes to $90^{\circ}$, the
survival probability $P_{ee}$ becomes 1, because the traveling path
length $L$ of the neutrinos approaches 0.

The results and the applications presented here are relevant to both the
atmospheric neutrino experiments ({\it e.g.} Super-Kamiokande) and the
LBL neutrino experiments. The present treatment assumes that in the
results of the atmospheric and LBL experiments, the scattering and
absorption of the neutrinos in matter is taken care of in the Monte
Carlo programs that are used to analyze the data from these
experiments. In a realistic analysis, we should also introduce damping
due to uncertainties in the energy $E_{\nu}$ of the neutrinos and
their path length $L$.

Examining some of the future LBL neutrino experiments, we find their
nadir angles to be $h \simeq 86.7^{\circ}$, $h \simeq 86.7^{\circ}$, and
$h \simeq 88.9^{\circ}$ for CERN-NGS ($L=743$ km), MINOS ($L=732$ km),
and K2K ($L=252$ km), respectively \cite{barg99,LBL}. See
Fig.~\ref{fig:4} for how the survival probability $P_{ee}$ depends on
the neutrino energy for these fixed nadir angles. Note that for these
three experiments the neutrinos only traverse the mantle and not the core.

There are, at the moment, no planned LBL experiments in which the
neutrinos also traverse the core. The baseline has in that case to be
longer than $L \simeq 10670$ km.

\ack

One of us (T.O.) would like to thank Martin Freund for useful discussions.
This work was supported by the Swedish Natural Science Research
Council (NFR), Contract No. F-AA/FU03281-312. Support for this work
was also provided by the Engineer Ernst Johnson Foundation (T.O.).

\newpage

\begin{figure}
\begin{center}
\epsfig{figure=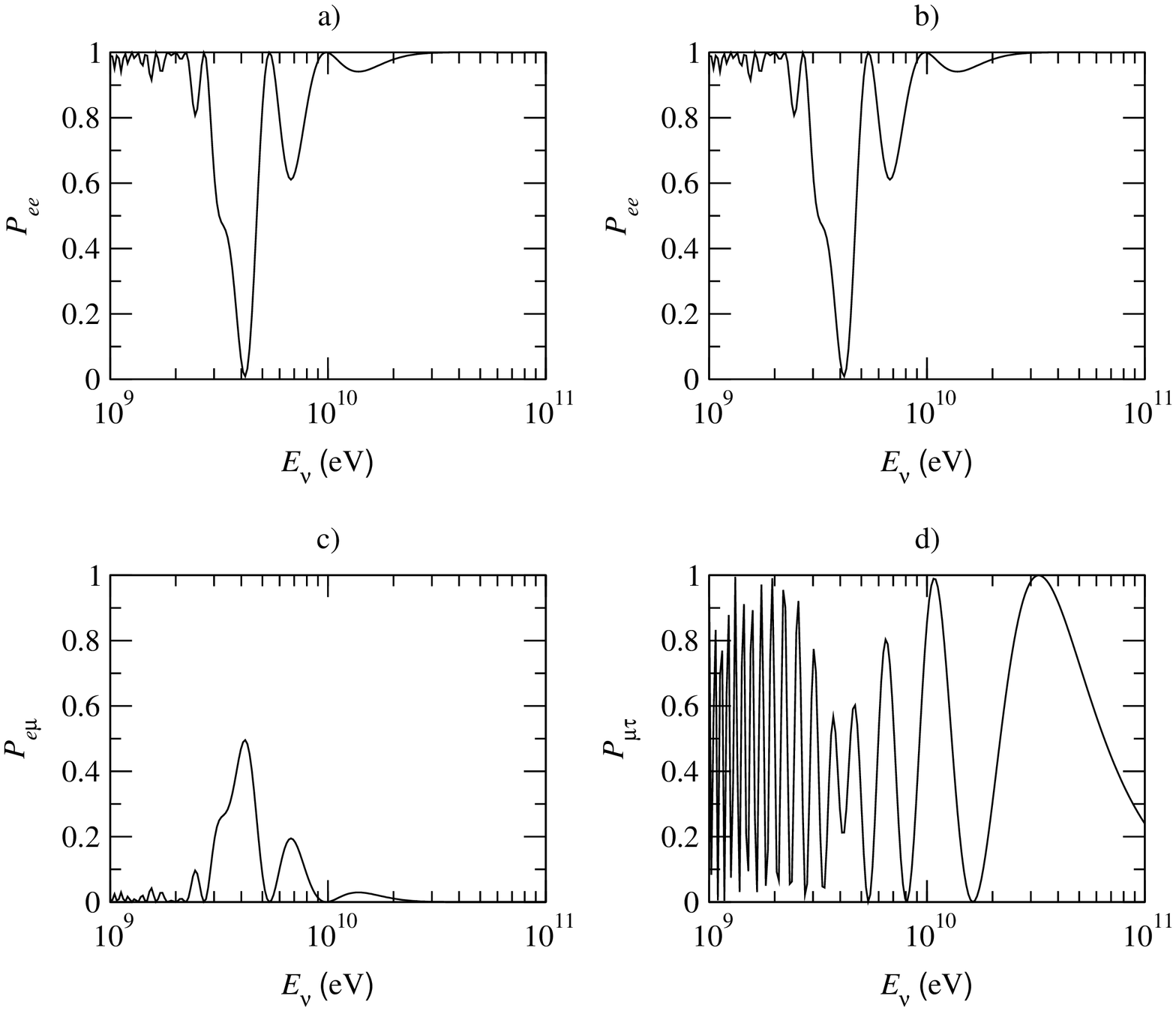,height=9cm,angle=-90,clip=on}
\caption{Transition probabilities as functions of the neutrino energy
$E_\nu$ for the parameter values $h=0$, $\theta_1 = \theta_3 = 45^\circ$
(bimaximal mixing), $\theta_2 = 5^\circ$, $\Delta M^2 = 3.2 \cdot
10^{-3}$ ${\rm eV}^2$, and $\Delta m^2 = 0$. a) $P_{ee}$ (two flavors;
parameters values: $\theta = 5^\circ$ and $\Delta m^2 = 3.2 \cdot 10^{-3}$
${\rm eV^2}$), b) $P_{ee}$ (three
flavors), c) $P_{e\mu}$ (three flavors), and d) $P_{\mu\tau}$ (three flavors).}
\label{fig:1}
\end{center}
\end{figure}

\begin{figure}
\begin{center}
\epsfig{figure=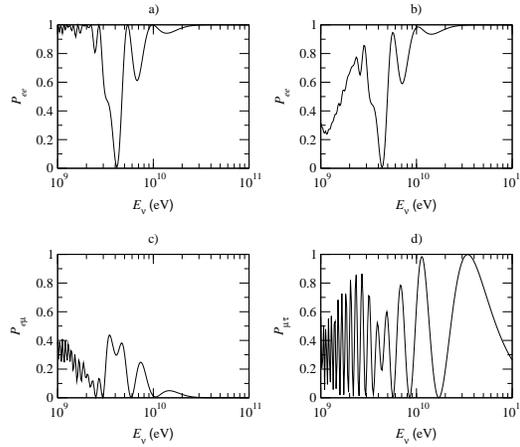,height=9cm,angle=-90,clip=on}
\caption{Transition probabilities as functions of the neutrino energy
$E_\nu$, using the same parameter values as for Fig.~\ref{fig:1}
except that $\Delta m^2 = \Delta M^2 / 10$. a) $P_{ee}$ (two flavors;
parameters values: $\theta = 5^\circ$ and $\Delta m^2 = 3.2 \cdot 10^{-3}$
${\rm eV^2}$),
b) $P_{ee}$ (three flavors), c) $P_{e\mu}$ (three flavors), and d)
$P_{\mu\tau}$ (three flavors).}
\label{fig:2}
\end{center}
\end{figure}

\begin{figure}
\begin{center}
\epsfig{figure=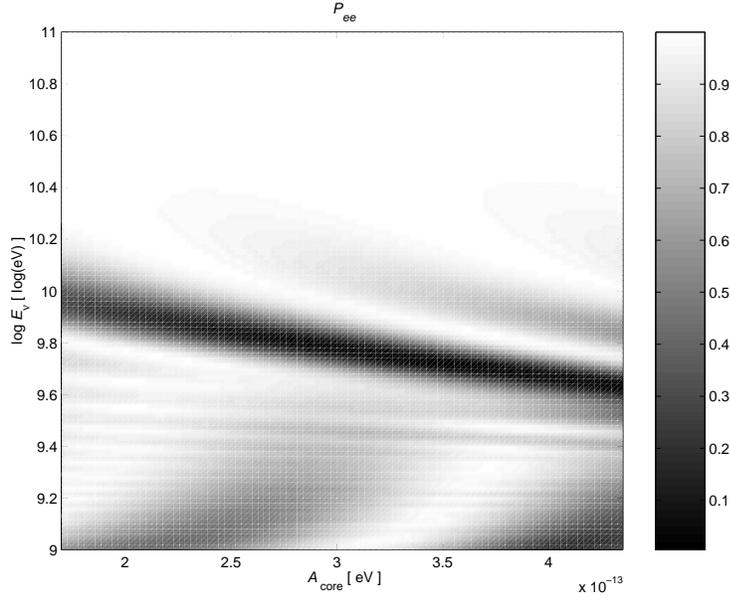,height=8cm,clip=on}
\caption{The survival probability $P_{ee}$ as a function of the core matter
density parameter $A_{\rm core} \in [A_1,A_2]$ and the
neutrino energy $E_\nu$, using the same parameter values as for
Fig.~\ref{fig:2}.}
\label{fig:3}
\end{center}
\end{figure}

\begin{figure}
\begin{center}
\epsfig{figure=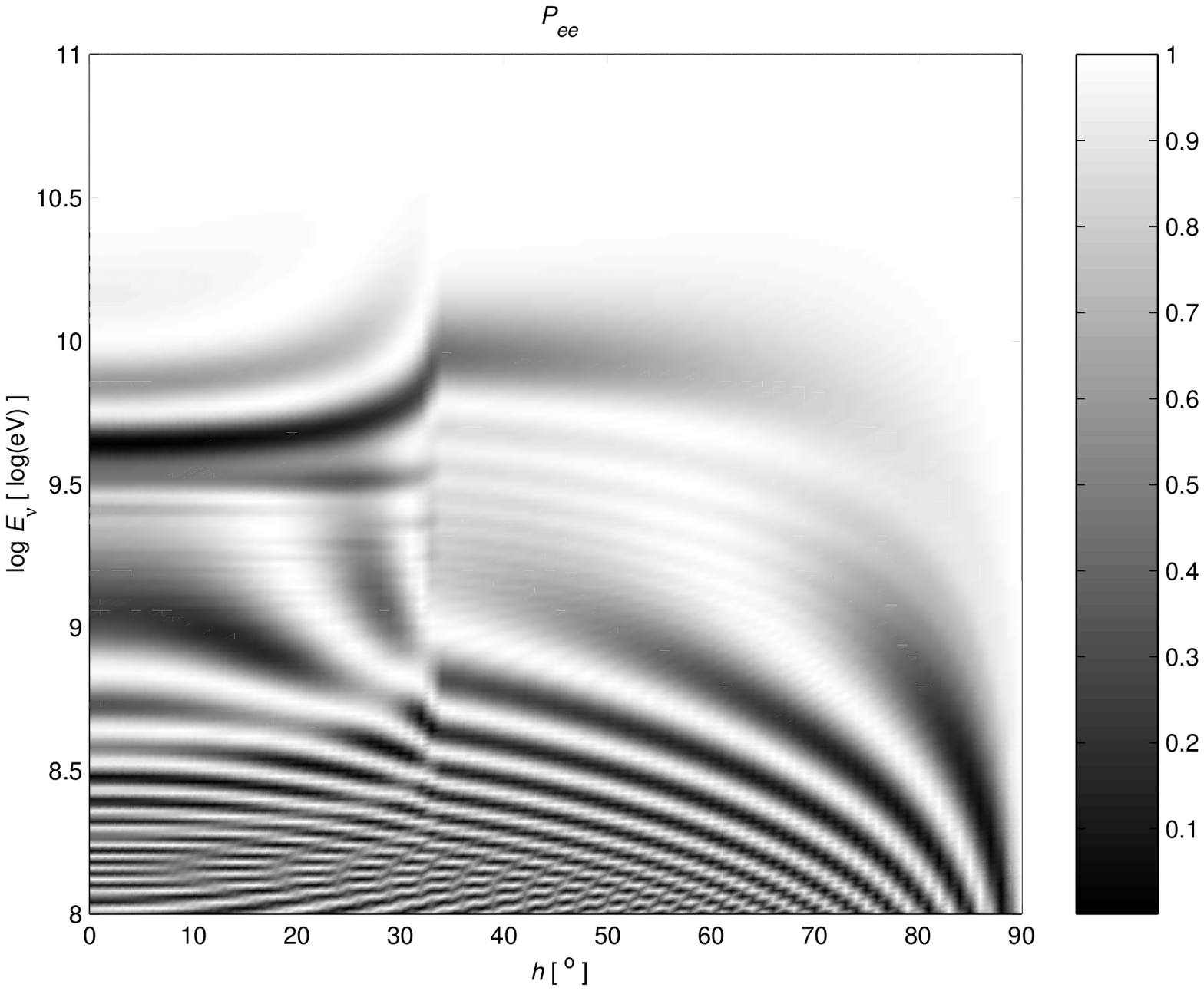,height=8cm,clip=on}
\caption{The survival probability $P_{ee}$ as a function of the nadir
angle $h$ and the neutrino energy $E_\nu$, using the same
parameter values as for Fig.~\ref{fig:2}.}
\label{fig:4}
\end{center}
\end{figure}


\begin{thebibliography}{9}

\bibitem{mikh85} S.P. Mikheyev and A.Yu. Smirnov,
Yad. Fiz. 42 (1985) 1441 [Sov. J. Nucl. Phys. 42 (1985)
913]; Nuovo Cimento C 9 (1986) 17.

\bibitem{wolf78} L. Wolfenstein, Phys. Rev. D 17 (1978) 2369;
20 (1979) 2634.

\bibitem{dick99} K. Dick, M. Freund, M. Lindner, and A. Romanino,
Nucl. Phys. B (to be published), hep-ph/9903308.

\bibitem{barg80} V. Barger, K. Whisnant, S. Pakvasa, and
R.J.N. Phillips, Phys. Rev. D 22 (1980) 2718.

\bibitem{kim87} C.W. Kim and W.K. Sze, Phys. Rev. D 35 (1987) 1404.

\bibitem{zagl88} H.W. Zaglauer and K.H. Schwarzer, Z. Phys. C
40 (1988) 273.

\bibitem{kuo86} T.K. Kuo and J. Pantaleone,
Phys. Rev. Lett. 57 (1986) 1805.

\bibitem{josh88} A.S. Joshipura and M.V.N. Murthy, Phys. Rev. D
37 (1988) 1374.

\bibitem{tosh87} S. Toshev, Phys. Lett. B 185 (1987) 177;
192 (1987) 478(E); S.T. Petcov and S. Toshev, Phys. Lett. B
187 (1987) 120; S.T. Petcov, Phys. Lett. B 214 (1988) 259.

\bibitem{oliv96} J.C. D'Olivo and J.A. Oteo, Phys. Rev. D 54 (1996)
1187.

\bibitem{fogl94} G.L. Fogli, E. Lisi, and D. Montanino, Phys. Rev. D
49 (1994) 3626; G.L. Fogli, E. Lisi, and D. Montanino,
Phys. Rev. D 54 (1996) 2048, hep-ph/9605273; G.L. Fogli,
E. Lisi, D. Montanino, and G. Scioscia, Phys. Rev. D 55
(1997) 4385, hep-ph/9607251; G.L. Fogli, E. Lisi, A. Marrone, and
G. Scioscia, Phys. Rev. D 59 (1999) 033001, hep-ph/9808205.

\bibitem{nico88} A. Nicolaidis, Phys. Lett. B 200 (1988) 553.

\bibitem{kras88} P.I. Krastev and S.T. Petcov, Phys. Lett. B
205 (1988) 84. 

\bibitem{giun98} C. Giunti, C.W. Kim, and M. Monteno, Nucl. Phys. B
521 (1998) 3, hep-ph/9709439.

\bibitem{liu98} Q.Y. Liu and A.Yu. Smirnov, Nucl. Phys. B
524 (1998) 505, hep-ph/9712493; Q.Y. Liu, S.P. Mikheyev, and
A.Yu. Smirnov, Phys. Lett. B 440 (1998) 319, hep-ph/9803415;
P. Lipari and M. Lusignoli, Phys. Rev. D 58 (1998) 073005,
hep-ph/9803440; E.Kh. Akhmedov, Nucl. Phys. B 538 (1999) 25,
hep-ph/9805272; E.Kh. Akhmedov, A. Dighe, P. Lipari, and
A.Yu. Smirnov, Nucl. Phys. B 542 (1999) 3, hep-ph/9808270;
E.Kh. Akhmedov, hep-ph/9903302; hep-ph/9907435.

\bibitem{petc98} S.T. Petcov, Phys. Lett. B 434 (1998) 321,
hep-ph/9805262; 444 (1998) 584(E); M. Chizhov, M. Maris, and
S.T. Petcov, hep-ph/9810501; M.V. Chizhov and S.T. Petcov,
Phys. Rev. Lett. 83 (1999) 1096, hep-ph/9903399; hep-ph/9903424.

\bibitem{harr99} P.F. Harrison, D.H. Perkins, and W.G. Scott,
Phys. Lett. B 458 (1999) 79.

\bibitem{freu99} M. Freund and T. Ohlsson, hep-ph/9909501.

\bibitem{kim93} C.W. Kim and A. Pevsner, {\it Neutrinos in Physics and
Astrophysics} (Harwood Academic, Chur, Switzerland, 1993).

\bibitem{caso98} Particle Data Group, C. Caso et al., {\it
Review of Particle Physics}, Eur. Phys. J. C 3 (1998) 1.

\bibitem{ohls99} T. Ohlsson and H. Snellman, J. Math. Phys. (to be
published), hep-ph/9910546.

\bibitem{scho99} K. Scholberg [Super-Kamiokande Collaboration], hep-ex/9905016.

\bibitem{bahc98} J.N. Bahcall, P.I. Krastev, and A.Yu. Smirnov,
Phys. Rev. D 58 (1998) 096016, hep-ph/9807216; 60 (1999)
093001, hep-ph/9905220.

\bibitem{apol98} CHOOZ Collaboration, M. Apollonio et al.,
Phys. Lett. B 420 (1998) 397, hep-ex/9711002; hep-ex/9907037.

\bibitem{barg99} V. Barger, S. Geer, R. Raja, and K. Whisnant, hep-ph/9911524.

\bibitem{LBL} CERN-LNGS Collaboration, P. Picchi and F. Pietropaolo,
Nucl. Phys. B (Proc. Suppl.) 77 (1999) 187; S.G. Wojcicki,
Nucl. Phys. B (Proc. Suppl.) 77 (1999) 182; K2K
Collaboration, K. Nishikawa, Nucl. Phys. B (Proc. Suppl.) 77 (1999)
198.

\end{thebibliography}
\end{document}